\documentclass[aps,prd,preprint,superscriptaddress,tightenlines,nofootinbib]{revtex4}

\usepackage{graphicx}
\usepackage{dcolumn}
\usepackage{bm}
\usepackage{epsfig}
\def\etal{{\em et al.}}

\newcommand{\cleoiii}{CLEO\,III}
\newcommand{\cleoc}{CLEO-c}
\newcommand{\ccb}{\mbox{$c\overline{c}$}}

\newcommand{\gev   }{\mbox{\rm GeV}}

\newcommand{\invpb }{\mbox{\rm pb$^{-1}$}}

\newcommand{\ks}{\mbox{$K_S$}}
\newcommand{\kl}{\mbox{$K_L$}}

\newcommand{\phot}{\mbox{$\gamma$}}
\newcommand{\epem}{\mbox{$e^+ e^-$}}

\newcommand{\Psip}{$\psi(2S)$}

\newcommand{\KsKl}{$K_{S}^{0} K_{L}^{0}$}
\newcommand{\toKK}{$\to K^{+} K^{-}$}
\newcommand{\pipi}{$\pi^{+}\pi^{-}$}
\newcommand{\topipi}{$\to \pi^{+}\pi^{-}$}
\newcommand{\KK}{$K^{+}K^{-}$}
\newcommand{\toKsKl}{$\to K_{S}^{0} K_{L}^{0}$}

\newcommand{\Ks}{$K_{S}^{0}$}
\newcommand{\Kl}{$K_{L}^{0}$}
\newcommand{\pio}{$\pi^{0}$}

\newcommand{\KKstar}{$K^{{\ast} 0}(892) \bar{K^0}$}
\newcommand{\chic}{$\chi_c$}
\begin{document}

\preprint{CLNS 05/1947}       
\preprint{CLEO 05-33}         

\title{Measurement of Interference between Electromagnetic and Strong Amplitudes in $\psi(2S)$ Decays to Two Pseudoscalar Mesons}

%
\author{S.~Dobbs}
\author{Z.~Metreveli}
\author{K.~K.~Seth}
\author{A.~Tomaradze}
\author{P.~Zweber}
\affiliation{Northwestern University, Evanston, Illinois 60208}
\author{J.~Ernst}
\affiliation{State University of New York at Albany, Albany, New York 12222}
\author{H.~Severini}
\affiliation{University of Oklahoma, Norman, Oklahoma 73019}
\author{S.~A.~Dytman}
\author{W.~Love}
\author{S.~Mehrabyan}
\author{V.~Savinov}
\affiliation{University of Pittsburgh, Pittsburgh, Pennsylvania 15260}
\author{O.~Aquines}
\author{Z.~Li}
\author{A.~Lopez}
\author{H.~Mendez}
\author{J.~Ramirez}
\affiliation{University of Puerto Rico, Mayaguez, Puerto Rico 00681}
\author{G.~S.~Huang}
\author{D.~H.~Miller}
\author{V.~Pavlunin}
\author{B.~Sanghi}
\author{I.~P.~J.~Shipsey}
\author{B.~Xin}
\affiliation{Purdue University, West Lafayette, Indiana 47907}
\author{G.~S.~Adams}
\author{M.~Anderson}
\author{J.~P.~Cummings}
\author{I.~Danko}
\author{J.~Napolitano}
\affiliation{Rensselaer Polytechnic Institute, Troy, New York 12180}
\author{Q.~He}
\author{J.~Insler}
\author{H.~Muramatsu}
\author{C.~S.~Park}
\author{E.~H.~Thorndike}
\affiliation{University of Rochester, Rochester, New York 14627}
\author{T.~E.~Coan}
\author{Y.~S.~Gao}
\author{F.~Liu}
\author{R.~Stroynowski}
\affiliation{Southern Methodist University, Dallas, Texas 75275}
\author{M.~Artuso}
\author{S.~Blusk}
\author{J.~Butt}
\author{J.~Li}
\author{N.~Menaa}
\author{R.~Mountain}
\author{S.~Nisar}
\author{K.~Randrianarivony}
\author{R.~Redjimi}
\author{R.~Sia}
\author{T.~Skwarnicki}
\author{S.~Stone}
\author{J.~C.~Wang}
\author{K.~Zhang}
\affiliation{Syracuse University, Syracuse, New York 13244}
\author{S.~E.~Csorna}
\affiliation{Vanderbilt University, Nashville, Tennessee 37235}
\author{G.~Bonvicini}
\author{D.~Cinabro}
\author{M.~Dubrovin}
\author{A.~Lincoln}
\affiliation{Wayne State University, Detroit, Michigan 48202}
\author{D.~M.~Asner}
\author{K.~W.~Edwards}
\affiliation{Carleton University, Ottawa, Ontario, Canada K1S 5B6} 
\author{R.~A.~Briere}
\author{I.~Brock}~\altaffiliation{Current address: Universit\"at Bonn, Nussallee 12, D-53115 Bonn}
\author{J.~Chen}
\author{T.~Ferguson}
\author{G.~Tatishvili}
\author{H.~Vogel}
\author{M.~E.~Watkins}
\affiliation{Carnegie Mellon University, Pittsburgh, Pennsylvania 15213}
\author{J.~L.~Rosner}
\affiliation{Enrico Fermi Institute, University of
Chicago, Chicago, Illinois 60637}
\author{N.~E.~Adam}
\author{J.~P.~Alexander}
\author{K.~Berkelman}
\author{D.~G.~Cassel}
\author{J.~E.~Duboscq}
\author{K.~M.~Ecklund}
\author{R.~Ehrlich}
\author{L.~Fields}
\author{R.~S.~Galik}
\author{L.~Gibbons}
\author{R.~Gray}
\author{S.~W.~Gray}
\author{D.~L.~Hartill}
\author{B.~K.~Heltsley}
\author{D.~Hertz}
\author{C.~D.~Jones}
\author{J.~Kandaswamy}
\author{D.~L.~Kreinick}
\author{V.~E.~Kuznetsov}
\author{H.~Mahlke-Kr\"uger}
\author{T.~O.~Meyer}
\author{P.~U.~E.~Onyisi}
\author{J.~R.~Patterson}
\author{D.~Peterson}
\author{E.~A.~Phillips}
\author{J.~Pivarski}
\author{D.~Riley}
\author{A.~Ryd}
\author{A.~J.~Sadoff}
\author{H.~Schwarthoff}
\author{X.~Shi}
\author{S.~Stroiney}
\author{W.~M.~Sun}
\author{T.~Wilksen}
\author{M.~Weinberger}
\affiliation{Cornell University, Ithaca, New York 14853}
\author{S.~B.~Athar}
\author{P.~Avery}
\author{L.~Breva-Newell}
\author{R.~Patel}
\author{V.~Potlia}
\author{H.~Stoeck}
\author{J.~Yelton}
\affiliation{University of Florida, Gainesville, Florida 32611}
\author{P.~Rubin}
\affiliation{George Mason University, Fairfax, Virginia 22030}
\author{C.~Cawlfield}
\author{B.~I.~Eisenstein}
\author{I.~Karliner}
\author{D.~Kim}
\author{N.~Lowrey}
\author{P.~Naik}
\author{C.~Sedlack}
\author{M.~Selen}
\author{E.~J.~White}
\author{J.~Wiss}
\affiliation{University of Illinois, Urbana-Champaign, Illinois 61801}
\author{M.~R.~Shepherd}
\affiliation{Indiana University, Bloomington, Indiana 47405 }
\author{D.~Besson}
\affiliation{University of Kansas, Lawrence, Kansas 66045}
\author{T.~K.~Pedlar}
\affiliation{Luther College, Decorah, Iowa 52101}
\author{D.~Cronin-Hennessy}
\author{K.~Y.~Gao}
\author{D.~T.~Gong}
\author{J.~Hietala}
\author{Y.~Kubota}
\author{T.~Klein}
\author{B.~W.~Lang}
\author{R.~Poling}
\author{A.~W.~Scott}
\author{A.~Smith}
\affiliation{University of Minnesota, Minneapolis, Minnesota 55455}
\collaboration{CLEO Collaboration} 
\noaffiliation


\date{March 6, 2006}

\begin{abstract}
Using a sample of $3.08\times 10^6$ \Psip\ decays collected at $\sqrt{s} = 3.686$ GeV 
with the CLEO detector at the Cornell Electron Storage Ring, 
we have measured the branching fractions for \Psip\ decays to pseudoscalar pairs \pipi,
\KK\ and \KsKl. We obtain 
${\cal B}$(\Psip$\rightarrow$~\pipi) $<$ $2.1\times10^{-5}$ (90\% C.L.), 
${\cal B}$(\Psip$\rightarrow$~\KK) 
= $(6.3\pm0.6({\rm stat})\pm0.3({\rm syst}))\times10^{-5}$, 
and ${\cal B}$(\Psip$\rightarrow$~\KsKl) 
= $(5.8\pm0.8({\rm stat})\pm0.4({\rm syst}))\times10^{-5}$. 
The branching fractions allow extraction of the relative phase $\Delta=(95\pm15)^{\circ}$ 
and strength ratio $R=(2.5\pm0.4)$ of the three-gluon and one-photon amplitudes 
for these modes.
\end{abstract}

\pacs{13.25.Gv, 12.38.Qk, 14.40.Gx}
\maketitle

The decay of narrow vector states of charmonium, $J/\psi$ and \Psip, into states of 
light quarks can proceed via \ccb\ annihilation into a virtual photon or three gluons. 
The decays of $J/\psi$ to pseudoscalar pairs (PP) have been analysed by several authors 
\cite{susuki previous,ros,gerard,susuki another reference}, and it has been determined 
that the decay $J/\psi$ \topipi\ proceeds dominantly through one photon, the decay $J/\psi$ 
\toKsKl\ proceeds dominantly through three gluons, and the decay $J/\psi$ \toKK\ may proceed 
through both one-photon and three-gluon channels, with a phase difference of nearly 
90$^{\circ}$ between the two amplitudes. Under the 
simplifying assumption that the SU(3) breaking correction to
the one-photon annihilation amplitude is negligibly small, Suzuki \cite{susuki previous} has determined the relative phase 
angle to be $\Delta(J/\psi)$ = (89.6 $\pm$ 9.9)$^{\circ}$ and Rosner \cite{ros} has 
confirmed it with the result $\Delta(J/\psi)$ = (89 $\pm$ 10)$^{\circ}$.  The 
branching fractions for the corresponding decays of \Psip\ were not available to either 
Gerard and Weyers \cite{gerard} or Suzuki \cite{susuki another reference} and both 
of them have speculated that although it may be naively expected that \Psip\ 
decays would have the same characteristics as $J/\psi$, it is possible that \Psip\ 
decays differ from those of $J/\psi$ in important ways. In particular, the available 
phase information obtained from \Psip\ decays to vector-pseudoscalar (VP) modes could 
not rule out the possibility that the
phase is near $180^{\circ}$ as suggested in Ref.~\cite{susuki another reference}.
In a recent attempt, Yuan \etal\ \cite{wang} used the 
unpublished BES results for ${\cal B}_{\pi^+\pi^-} \equiv {\cal B}$(\Psip\topipi) and 
${\cal B}_{K^+ K^-}\equiv {\cal B}$(\Psip\toKK) to study the interference between 
electromagnetic and strong amplitudes, 
but were not able to obtain an estimate of their relative phase and magnitude in the 
absence of a measurement of the branching fraction for the 
decay \Psip\ \toKsKl. A subsequent measurement by BES of 
${\cal B}_{K_s^0K_L^0}\equiv {\cal B}$(\Psip\toKsKl)~\cite{BES psi2s paper}, 
together with the older BES results for the other two decays, has led to 
$\Delta$(\Psip) = ($-$82 $\pm$ 29)$^{\circ}$ or (+121 $\pm$ 27)$^{\circ}$.  

In this Letter, we report on the results of CLEO measurements of all
the three branching fractions. We have an improved \KK\ branching
fraction measurement and the \KsKl\ mode is 
essentially background free and consistent with the earlier BES
measurement. Using our measurements, we determine the ratio of the 
amplitudes of \Psip $\to$ PP decays via a photon and three gluons and the 
phase difference between the two. We also report on the ratios of 
${\cal B}$(\Psip)/${\cal B}(J/\psi)$ for the three decays to test the 
``12$\%$'' rule \cite{rule}.

The data used in this analysis were collected at the CESR \epem\ storage ring,
which has been reconfigured to run in the charm meson region by insertion of 
wiggler magnets~\cite{cleoc}.
Our analysis is based on 
3.08$\times 10^6$ $\psi(2S)$ 
decays, which corresponds to a total integrated luminosity of 5.63~\invpb. 
Approximately half of these data (2.74~\invpb) 
were taken with the \cleoiii\ detector configuration~\cite{cleo}, 
while the remainder (2.89~\invpb) of the $\psi(2S)$ data 
together with 20.7~pb$^{-1}$ of off-resonance data 
taken at $\sqrt{s} = 3.671~\gev$ 
were collected with the reconfigured \cleoc\ detector~\cite{cleoc}.
Both detector configurations are cylindrically symmetric and provide 
93$\%$ coverage of solid angle for charged and neutral particle 
indentification.  
The detector components important for this analysis are the main drift chamber (DR), 
the Ring-Imaging CHerenkov detector (RICH), and the CsI crystal calorimeter (CC), 
all of which are common to both detector configurations.   

The properties of the PP modes are studied by generating
Monte Carlo events ({\sc evtgen} event generator~\cite{evtgen}) using simulations 
of each of the two detector
configurations and for each of the three decay modes using a 
{\sc geant}-based \cite{geant} detector modeling program.
For all three modes, events are simulated with $\sin^2\theta$ angular distributions, 
where $\theta$ is the angle between the decay product and the positron beam in the 
center-of-mass system, as is expected for a vector resonance decaying into 
two pseudoscalar mesons.

The events for each of the charged PP decay modes are required to have 
two charged particles and zero net charge. In the case of the neutral
PP mode, \Ks\ candidates are formed from a pair of two charged tracks
which are constrained to come from a common vertex, are consistent with the pion hypothesis, 
and possess an invariant mass within 10~MeV ($\approx 3.2\sigma$) of the 
nominal \Ks\ mass.  The charged particles 
in the  \pipi, \KK, and \KsKl\ (i.e., the \pipi\ daughters of the \Ks) 
decay modes are required to have $|$cos$\theta|$ $<$~0.75, $<$~0.93 and $<$~0.93,  
respectively. In the case of the \pipi\ mode, the additional 
requirement of an associated shower in the CC is imposed.  
Furthermore, each of the charged particles is required to satisfy 
standard criteria for track quality and distance of closest
approach to the interaction point.  For the neutral PP mode  the
latter requirement is reversed and a displaced secondary \Ks\ vertex
(intersection of the \pipi\ daughters) condition of $>$ 5 mm is imposed. 

We require momentum conservation in the reconstructed 
charged PP events by demanding the vector sum of the total momentum  
in an event, $| \Sigma \vec{\bf p}| / E_{\rm beam}$, 
be $<$~0.04 for the \KK\ mode and $<$~0.054 for the \pipi\ mode. 
This eliminates background via $\pi\pi J/\psi$ or $\phot\chi_{\rm cJ}$
($J=$~0,~1~and~2) decays of the \Psip.

To optimize the discrimination between $p$, $K$, $\pi$, $\mu$ and $e$, 
we combine the particle identification information obtained 
from the specific ionization ($dE/dx$) measured in the DR with that obtained 
from the RICH detector to form a joint $\chi^2$ function. For $dE/dx$, 
we form a quantity $S_i$ ($i=p,K,\pi,\mu,~{\rm and}~e$), which is the
difference between the measured and expected $dE/dx$ for that
hypothesis, normalized to its standard deviation.
The information from the RICH is given in the form of a likelihood function, 
$-2{\rm log}L$. The joint $\chi^2$ function is
$\Delta\chi^2 (i-j)= -2{\rm log}L_i + 2{\rm log}L_j + S_i^2 - S_j^2$. 
The more negative $\Delta\chi^2$, the higher the likelihood that
the particle is of type~$i$ compared to type~$j$. The requirement on 
these quantities varies in value from mode to mode depending upon background 
considerations. For kaons in the \KK\ decay mode for 3$\sigma$ separation we require 
$\Delta\chi^2(K-p) < -9 $ and $\Delta\chi^2(K-\pi) < -9$. 
For pions in the \pipi\ decay mode 
we require looser particle identification criteria 
$\Delta\chi^2(\pi-e) < 0$ and $\Delta\chi^2(\pi-K) < 0$. In the case
of \pipi\ daughters of the \Ks\ in the neutral \KsKl\ mode, we impose
similar criteria $\Delta\chi^2(\pi-K) < 0$ and $\Delta\chi^2(\pi-p) < 0$.

QED processes ($e^{+}e^{-} \to \gamma \gamma$ and $l^{+}l^{-}$) 
are possible background sources in these final states. 
To combat the charged di-lepton contamination, 
in the case of the \KK\ mode, we require $\Delta\chi^2(K-e/\mu) < -9$ 
for the charged tracks.  
For the \KsKl\ mode, we reject as an
electron any daughter pion track with the ratio of CC energy
$E_{\mathrm{CC}}$ to track momentum of $p$ 0.92~$<~E_{\mathrm{CC}}/p<$~1.05 
and $\Delta\chi^2(\pi-e) < 0$.  
In order to suppress the more severe $l^+l^-$ background events in the \pipi\
decay mode, it is required that the 
accepted events have $E_{\mathrm{CC}}/p <$~0.85 for each candidate pion 
track. Rejecting $\mu^+\mu^-$ background events from the \pipi\ sample 
requires additional measures, which 
are determined by studying $\mu$ tracks from a 
$e^+e^-\to\mu^+\mu^-$ simulation sample and by studying pions of appropriate 
momenta ($\sim1.83$ GeV/$c$) in the existing CLEO sample 
of inclusive $D^0\to K^-\pi^+$ data taken at $\sqrt{s}=10.58$~GeV.  The 
optimization criteria which emerge from these studies are that pions must 
deposit $E_{\mathrm{CC}}>0.42$~GeV, and the average efficiency from simulations and
data pion track samples is applied for each track.
For the \KsKl\ mode, we do not attempt to reconstruct the \Kl, and so cannot
reconstruct the complete event.  
In order to reject most of the anticipated backgrounds 
(discussed later), we designed the following selection criteria to
reject events with neutral particles other than a \Kl\ 
accompanying the reconstructed \Ks\ in an event. 
We require the energy of the shower associated with neutrals, and closest to the 
inferred \Kl\ direction (obtained from the 4-momentum of the beam 
and the reconstructed \Ks), to be less than 1.5~GeV.  
We define a cone of 0.35~radians around the \Kl\ direction and require
that of all showers associated with neutrals outside (inside) this cone, there be 
none (at most one) with $E>100$~MeV, and that the sum of all showers outside this 
cone does not exceed 300~MeV.
We also have an explicit \pio\ veto for even
better rejection of events which have one or more \pio\ mesons from
neutral sources of backgrounds. 
Simulation studies show that after these cuts, 
there is minimal QED background contamination
in the \KsKl\ final state.

For each charged PP candidate event, we calculate the scaled
visible energy, $E_{\rm vis}/\sqrt{s}$, where $E_{\rm vis}$ is the energy reconstructed 
in an event and $\sqrt{s}$ is the center-of-mass energy. 
For each \KsKl\ candidate event, we calculate the scaled \Ks\ energy, 
$E_{K_{S}^{0}}/E_{\rm beam}$, where $E_{K_{S}^{0}}$ is the measured \Ks\ energy and 
$E_{\rm beam}$ is the beam energy.  We define our signal region to be 
0.98~$<$ $E_{\rm vis}/\sqrt{s}$ (or $E_{K_{S}^{0}}/E_{\rm beam}$) $<$~1.02, 
and two sideband regions of 0.94-0.98 and 1.02-1.06, as representative of the 
combinatorial background. 

We also study the data sample taken at $\sqrt{s} = 3.671$ GeV (continuum) to check for 
possible non-resonant contributions in our $\psi(2S)$ signals. This is 
found to be non-negligible for the charged PP decay modes 
as shown in Table~\ref{tab:result}. We multiply the 
yield from the continuum data by a scaling factor which is calculated 
taking into account the luminosity ratio (5.63/20.7~=~0.272), 
a $1/s^3$ correction for mesons, and the values of the efficiencies in the \cleoiii\ and
\cleoc\ detector configurations before subtracting it from the
$\psi(2S)$ yields. 
The scale factors are $f_{s}$ =~0.265, 0.261 and 0.250 for the 
\pipi, \KK\ and \KsKl\ final states, respectively.

\begin{figure*}
\includegraphics*[width=6.5in]{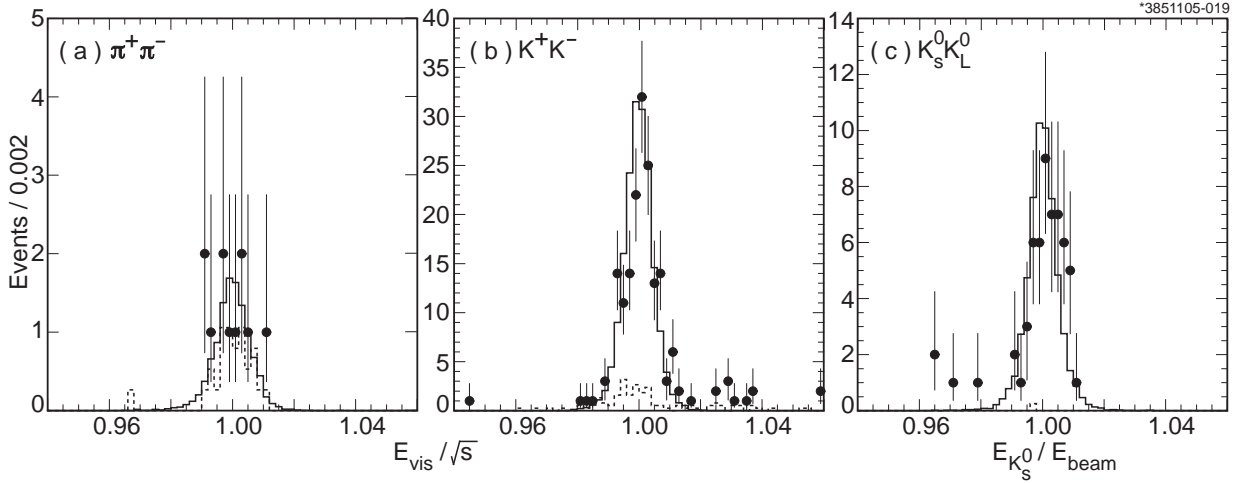}
\caption{Scaled energy ($E_{\rm vis}/\sqrt{s}$) distributions for the 
$\psi(2S)$ $\rightarrow$ (a) \pipi\  and (b) \KK\ decay modes and 
the scaled $K^{0}_{s}$ energy ($E_{K_{S}^{0}}/E_{{\rm beam}}$) distribution for the 
$\psi(2S)$ $\rightarrow$ (c) \kl\ks\ decay mode.  
Signal data, signal simulations, and scaled non-resonant data are shown 
as points, solid histograms, and dashed histograms, respectively.  The signal simulations 
are normalized to the number of observed events in their respective signal regions.}
\label{fig:data}
\end{figure*}

Figure~\ref{fig:data} shows the scaled energy distribution for each of the 
decay modes with the points showing the data, the solid histogram showing the simulation 
results, and the dashed histogram showing the scaled non-resonant contribution.  
In all modes clear signals are seen, with widths consistent 
with those expected from simulation studies. 
The \pipi\ mode, shown in Figure~\ref{fig:data}(a), 
is statistics limited and has no
combinatorial background, but there is a large background
contribution from non-resonant $e^+e^-$ annihilation events. 
Figure~\ref{fig:data}(b) for the \KK\ mode shows an excess due to 
mis-identified dilepton pairs around $E_{\rm vis}/\sqrt{s}$~=~1.03 (high sideband region), 
and some non-negligible non-resonant background is also present. 
In the \KsKl\ mode, shown in Figure~\ref{fig:data}(c), the
background is asymmetric. The low sideband region around
$E_{K_{S}^{0}}/E_{\rm beam} = 0.96$ is contaminated from known hadronic sources
such as \KKstar $+\mathrm{c.c.}$ (both the \Psip~\cite{VP paper of Hanna} and the continuum below the resonance can produce this final state), and the $K_{S}^{0}K_{S}^{0}$ final state which is possible through radiative transitions to
the \chic$_{0,2}$~\cite{PDG 2004}. We account
for possible background contamination inside the signal region by a
systematic uncertainty component obtained from simulation studies.

In the \pipi\ and \KK\ modes the signal yield, $N_S$, is obtained by subtracting 
from the observed yield, $S_{\psi(2S)}$, the QED (scaled sideband) contribution, $N_{\mathrm{QED}}$, and the 
scaled contribution, $f_s\cdot N_{\mathrm {cont}}$, from the observed yield in the continuum (minus the QED contribution in it), $N_{\mathrm {cont}}$. A possible contamination from the \Psip\ tail in the continuum yield is found to be negligibly small in all the modes.   
In other words, 
$N_S = S_{\psi(2S)}-N_{\mathrm{QED}}-f_s\cdot N_{\mathrm {cont}}$.  
In the \KsKl\ mode, as stated earlier, there is no QED background.  However, one count 
was observed in the continuum data, leading to $f_s\cdot N_{\mathrm {cont}}$ =~0.3. In Table \ref{tab:result} the observed yields and the 
subtractions are listed, as are the efficiencies calculated using the luminosity weighted average from the simulation studies for the CLEO III and CLEO-c detectors.  The 
branching fractions are obtained as $N_S/[N(\psi(2S))\cdot\epsilon]$, with 
$N(\psi(2S))$ =~3.08$\times10^6$.  For \pipi\ both the measured value and the 
corresponding 90$\%$ confidence upper limit are listed.  
 For \KsKl\ the listed value is obtained using ${\cal B}$(\Ks\ \topipi) 
= (68.95$\pm$0.14)$\%$ \cite{PDG 2004}.  
The entries for $Q$ are obtained using the literature \cite{PDG 2004,J/psi numbers for Q} 
values for $J/\psi$ branching fractions with statistical and systematic errors 
combined in quadrature.  

We evaluate the following systematic uncertainties to our measured
branching fractions for the charged (neutral) PP modes:
3.0\% uncertainty on the number of \Psip\ decays in our sample; 
1.0\% (2.0\%) uncertainty in the simulation of our hardware trigger; 
1.0\% (1.4\%) uncertainty in the reconstruction of each charged track in the event; 
1.0\% and 2.7\% (0.6\%) uncertainties for kaon and pion identification.  
Uncertainties in the \pipi, \KK, and \KsKl\ modes arising due to background 
subtraction procedures are 0.2\%, 1.4\%, and 3.8\%; and those due to simulation 
statistics are 1.3\%, 1.6\%, and 0.4\% respectively.
Additional uncertainties in the \pipi mode from the 
$E_{\rm vis}/\sqrt{s}$, $| \Sigma \vec{\bf p}| / E_{\rm beam}$, and 
$E_{\mathrm{CC}}$ criteria are determined to be 1.4\%, 24.3\%, and 12.1\%,
respectively. In the \KsKl\ mode, additional uncertainties arise due to
\Kl\ selection (3.7\%), \Ks\ finding (3.0\%), and
the $\cal{B}$(\Ks\ $\to \pi^{+}\pi^{-} $)~\cite{PDG 2004} (0.1\%). 
After combining all contributions in quadrature, the total systematic 
uncertainties for the \pipi, \KK, and \KsKl\ final states are 
28.0\%, 4.7\% and 7.3\%, respectively. 

\begin{small}
\begin{table*}[thp]
\begin{center}
\caption{Experimental results for \Psip\ decays into pairs of pseudoscalar mesons.  The 
first errors are statistical and the second errors are systematic.  The last column shows 
the ``$Q$'' value which is the ratio of ${\cal B}(\psi(2S) \to PP)$ to ${\cal B}(J/\psi \to PP)$~\cite{PDG 2004,J/psi numbers for Q} with statistical and systematic 
uncertainties added in quadrature.}
\label{tab:result}
\begin{tabular}{l c c c c c c c} \hline
Modes & ~~$S_{\psi(2S)}$ & ~~$N_{\mathrm{QED}}$ & ~~$f_S\cdot N_{\mathrm {cont}}$ & $N_S$ 
& ~~$\epsilon$($\%$) & {${\cal B}(10^{-5})$} & $Q$($\%$) \\ 
\hline
\pipi  & 11   & $<$0.1  &  6.8  & 4.2  & 16.7 & 0.8$\pm$0.8$\pm$0.2  & 5.4$\pm$5.6  \\
       &      &         &       &       &      & ~~~$<$ 2.1 (90$\%$ C.L.) & ~~~$<$ 14.6 (90$\%$ C.L.) \\
\KK    & 163  & 6.0   & 17.8  & 139.2 & 71.7 & 6.3$\pm$0.6$\pm$0.3  & ~~26.6$\pm$4.5 \\
\KsKl  & 53   & -     & 0.3   & 52.7  & 42.8 & 5.8$\pm$0.8$\pm$0.4  & ~~32.2$\pm$5.2 \\ 
\hline
\end{tabular}
\end{center}
\end{table*}
\end{small}

The results in  Table \ref{tab:result} show that the statistical errors are larger than the 
systematic errors, particularly for the \pipi\ mode.  
The branching fractions ${\cal B}_{K^+ K^-}$ and ${\cal B}_{K_s^0K_L^0}$ 
are larger than predicted by the ``12$\%$ rule''.

In Table \ref{tab:comparison} we summarize the branching ratios for the three PP decays 
from the literature, from our measurements, and the world averages.  We also list the 
resulting ratios of the strong/electromagnetic amplitudes and the phase differences 
between them.  The procedure for determining these is described in the following.  

\begin{small}
\begin{table}[thp]
\begin{center}
\caption{Branching fractions used for determination of strong/EM interference parameters.  
The statistical and systematic errors have been combined in quadrature.  The branching 
fractions are at the $10^{-5}$ level.}
\label{tab:comparison}
\begin{tabular}{c c c c c} \hline
& ~~~DASP \cite{DASP results} 
& ~~~BES \cite{BES psi2s paper,BES charged modes} & CLEO & World Avg. \\
\hline
${\cal B}_{\pi^+\pi^-}$  
& 8$\pm$5  & 0.84$\pm$0.65 & 0.8$\pm$0.8 & 0.9$\pm$0.5 \\
${\cal B}_{K^+ K^-}$    
& 10$\pm$7 & 6.1$\pm$2.1   & 6.3$\pm$0.7 & 6.3$\pm$0.7 \\
${\cal B}_{K_s^0K_L^0}$  
& ---      & 5.24$\pm$0.67 & 5.8$\pm$0.9 & 5.4$\pm$0.6 \\
\hline
$R(\psi(2S))$      & --- & 2.6$\pm$1.0 & 2.8$\pm$1.4 & 2.6$\pm$0.7 \\
$\Delta(\psi(2S))$ & --- & (89$\pm$35)$^{\circ}$ & (93$\pm$20)$^{\circ}$ 
& (89$\pm$14)$^{\circ}$ \\
\hline
\end{tabular}
\end{center}
\end{table}
\end{small}

It was noted by both Suzuki \cite{susuki previous} and Rosner \cite{ros} that the 
available data for $J/\psi$ did not have enough precision to take account of the small 
amplitude for the SU(3) breaking strong decay.  Neither was it possible to take 
account of the resonance interference with the continuum.  For \Psip\ decays the 
statistical precision is even poorer.  We are therefore obliged to also forego the 
consideration of the SU(3) breaking and \Psip-continuum interference.  With these 
assumptions, following Rosner \cite{ros}, we obtain the phase difference 
and ratio of amplitudes between the strong three-gluon, $A(ggg)$, 
and electromagnetic, $A(\gamma)$, decays as follows:

\begin{equation}
R(\psi(2S)) = \frac{A(ggg)}{A(\gamma)} = \sqrt{\frac{{\cal B}_{K_s^0K_L^0}}
{\rho{\cal B}_{\pi^+\pi^-}}},
\end{equation}

\begin{equation}
\Delta(\psi(2S)) = \mathrm{cos}^{-1}\left( \frac{{\cal B}_{K^+ K^-} 
- {\cal B}_{K_s^0K_L^0} - \rho{\cal B}_{\pi^+\pi^-}}
{2\sqrt{{\cal B}_{K_s^0K_L^0}\cdot\rho{\cal B}_{\pi^+\pi^-}}}\right),
\end{equation}
where the phase space ratio $\rho$ = $(p_K/p_\pi)^3$ = 0.902.

The results in Table \ref{tab:comparison} are further improved if the CLEO result 
for ${\cal B}_{\pi^+\pi^-}$ from direct counting is replaced by that obtained 
from the 
recent CLEO measurement of the charged pion form factor $|F_{\pi}(\sqrt{s}$ = 3.671~GeV$)|$ 
= 0.075 $\pm$ 0.009 \cite{tlff}.  Under the assumption that the \Psip\ decay to \pipi\ 
is purely electromagnetic, it can be shown that \cite{wang another ref} 
\begin{equation}
{\cal B}_{\pi^+\pi^-} = 2 {\cal B}_{e^+e^-}
\left(\frac{p_{\pi}}{M_{\psi(2S)}}\right)^3|F_{\pi}(M^2_{\psi(2S)})|^2.  
\end{equation}
Using this relation we obtain 
${\cal B}_{\pi^+\pi^-}$ = $(1.04\pm0.23)\times10^{-5}$.  With this value 
$R$(CLEO) = 2.5$\pm$0.4 and $\Delta$(CLEO) = (95$\pm$15)$^{\circ}$, and 
$R$(World Avg.) = 2.4$\pm$0.4 and $\Delta$(World Avg.) = (90$\pm$12)$^{\circ}$.

As a by-product of this work, from the one count for \KsKl\ observed in the 
continuum data, we obtain the 90$\%$ C.L. of $\sigma_0$(\KsKl) $<$ 0.74~pb 
at $\sqrt{s}$ = 3.671 GeV including radiative corrections
from the \KK\ analysis in Ref.~\cite{tlff}. Using the relation \cite{Petes ref 26} 
\begin{equation}
\sigma_0(s) = \frac{\pi\alpha^2}{3s}\beta_{K^0}^3|F_{K^0}(s)|^2,
\end{equation}
where $\alpha$ is the fine-structure constant, $\beta_{K^0}$ is the \Ks\ velocity in 
the laboratory system, and $|F_{K^0}(s)|$ is the neutral kaon electromagnetic form 
factor, we obtain $|F_{K^{0}}(s$~=~13.48~GeV$^2)| < $~0.023 at 90\% C.L. including 
systematic uncertainties obtained for the \KsKl\ mode. Previous 
measurements of this form factor were limited to 
$s <$~4.5~GeV$^2$ \cite{new form fact refs2}.

In conclusion, we have analyzed CLEO~III and CLEO-c $\psi(2S)$ 
data corresponding to 3.08$\times 10^6$ \Psip\ decays, and
 have presented new measurements of the branching fractions into \pipi, 
\KK\ and \KsKl\ final states.  This has allowed the determination of parameters of the 
interference between the amplitudes for the strong and electromagnetic decays of \Psip\ 
into pseudoscalar pairs.  In particular, the phase difference between the two amplitudes 
is found to be nearly 90$^{\circ}$, in contrast to some earlier theoretical speculations 
\cite{gerard,susuki another reference}.

We gratefully acknowledge the effort of the CESR staff 
in providing us with excellent luminosity and running conditions.
This work was supported by 
the A.P.~Sloan Foundation,
the National Science Foundation,
and the U.S. Department of Energy.

 

\begin{thebibliography}{99}

\bibitem{susuki previous}
M. Suzuki, Phys. Rev. D {\bf 60}, 051501(R) (1999).

\bibitem{ros}
J. L. Rosner, Phys. Rev. D {\bf 60}, 074029 (1999).

\bibitem{gerard}
J.-M. Gerard and J. Weyers, Phys. Lett. B {\bf 462}, 324 (1999).

\bibitem{susuki another reference}
M. Suzuki, Phys. Rev. D {\bf 63}, 054021 (2001).

\bibitem{wang}
C. Z. Yuan, P. Wang, and X. H. Mo, Phys. Lett. B {\bf 567}, 73 (2003).

\bibitem{BES psi2s paper}
BES Collaboration, J. Z. Bai \etal, Phys. Rev. Lett. {\bf 92}, 052001 (2004).

\bibitem{rule}
W. S. Hou and A. Soni, Phys. Rev. Lett. {\bf 50}, 569 (1983);
Mark\,II Collaboration, M. E. B. Franklin \etal, Phys. Rev. Lett. {\bf 51}, 963 (1983);
W. S. Hou, Phys. Rev. D {\bf 55}, 6952 (1997);
Y. F. Gu and X. H. Li, Phys. Rev. D {\bf 63}, 114019 (2001).


\bibitem{cleoc}
CESR-c and CLEO-c Taskforces, CLEO-c Collaboration, R. A. Briere \etal, 
Cornell University, LEPP Report No. CLNS 01/1742, (2001) (unpublished).

\bibitem{cleo}
 G. Viehhauser {\etal}, Nucl. Instrum. Methods A {\bf 462}, 146 (2001);
 D. Peterson {\etal}, Nucl. Instrum. Methods Phys. Res. A {\bf 478}, 142 (2002);
 M. Artuso {\etal}, Nucl. Instrum. Methods A {\bf 554}, 147 (2005).

\bibitem{evtgen}
D.J. Lange, Nucl. Instrum. Methods Phys. Res., Sect. A {\bf 462}, 152 (2001).

\bibitem{geant}
R. Brun \etal, {\textsc Geant} 3.21, CERN Program Library Long Writeup W5013, 1993 (unpublished).

\bibitem{VP paper of Hanna}
CLEO Collaboration, N. E. Adam \etal, Phys. Rev. Lett. {\bf 94}, 012005 (2005); 
CLEO Collaboration, R. A. Briere \etal, Phys. Rev. Lett. {\bf 95}, 062001 (2005).

\bibitem{PDG 2004}
Particle Data Group, S. Eidelman \etal,  Phys. Lett. B {\bf 592}, 1 (2004).

\bibitem{J/psi numbers for Q}
BES Collaboration, J. Z. Bai \etal, Phys. Rev. D {\bf 69}, 012003 (2004).


\bibitem{DASP results}
DASP Collaboration, R. Brandelik \etal, Z. Phys. C1, 233 (1979).

\bibitem{BES charged modes}
S. W. Ye, ``Study of some VP and PP modes of \Psip\ decays'',
Ph.D. thesis, University of Science and Technology of China, 1997 (in Chinese).

\bibitem{tlff}
CLEO Collaboration, T. K. Pedlar \etal, Phys. Rev. Lett. {\bf 95}, 261803 (2005).

\bibitem{wang another ref}
P. Wang, C. Z. Yuan and X. H. Mo, Phys. Rev. D {\bf 69}, 057502 (2004).


\bibitem{Petes ref 26}
N. Cabibbo and R. Gatto, Phys. Rev. {\bf 124}, 1577 (1961).

\bibitem{new form fact refs2}
DM1 Collaboration, F. Mane \etal, Phys. Lett. B {\bf 99}, 261 (1981).











\end{thebibliography}
\end{document}